\documentclass[aps,prl,amsmath,amssymb,twocolumn,showpacs]{revtex4}
\usepackage{amssymb}
\usepackage{graphicx}

\begin{document}

\title{Routes to gelation in a clay suspension}
\author{B.~Ruzicka$^1$, L.~Zulian$^1$, G.~Ruocco$^{1,2}$}
\affiliation{ $^1$ INFM and Dipartimento di Fisica, Universit\'a
di Roma "La Sapienza", P.zle A. Moro 2, I-00185 Roma, Italy}
\affiliation{$^2$ INFM-CRS SOFT, Universit\'a di Roma "La
Sapienza", P.zle A. Moro 2, I-00185, Roma, Italy.}
\date{\today}
\begin{abstract}
The gelation of water suspension of a synthetic clay (Laponite)
has been studied by dynamic light scattering in a wide range of
clay weight concentration ($C_w=0.003 \div 0.031$). At variance
with previous determination, indicating a stable liquid phase for
$C_w<C_w^* \approx 0.015 \div 0.018$, we find that the gelation
takes actually place in the whole examined $C_w$ range. More
importantly, we find that $C_w^*$ marks the transition between two
different routes to gelation. We hypothesize that at low
concentration Laponite suspension behaves as an attractive colloid
and that the slowing down of the dynamics is attained by the
formation of larger and larger clusters while at high
concentration the basic units of the arrested phase could be the
Debye Huckel spheres associated to single Laponite plates.

\end{abstract}
\pacs{ 64.70.Pf, 82.70.Dd, 78.35.+c, 61.20.Lc} \maketitle

Colloidal systems are ideal benchmarks for studying, by optical
microscopy and light scattering, equilibrium slow dynamical
processes and the formation of non ergodic arrested state of
matter.  In this context sterically stabilized colloidal systems
have been often used as model for hard spheres, to study the
fluid-crystal transition and the glass formation\cite{Pusey}. For
suspensions of hard spheres the latter transition occurs at
relatively high packing fraction ($\varphi\geq 0.58$). When
screened charged interactions are present, crystal and disorder
arrested phases can be formed also at very low volume fraction
($\varphi\approx 0.01$). Recent works \cite{Teoria1,
Teoria2,Esperimenti1, Esperimenti2} focusing on short-range
attractive colloids have attempted a connection between the gel
and the glass arrested state of matter. In these systems, a
re-entrant glass line, two kind of glasses (attractive and
repulsive) and a glass-glass line have been predicted
\cite{Teoria1, Teoria2, Teoria3} and experimentally observed
\cite{Esperimenti1, Esperimenti2}. Correlations between the
dynamical behavior of gels and glasses suggest that a common
understanding of these two disordered forms of matter may emerge.
In the case further complicated where short-ranged attractive
interactions are complemented by weak repulsive electrostatic
interactions, the gel formation process can be fully modeled as a
glass transition phenomenon \cite{Sciortino}. In the last case,
the gel phase is stabilized by the competition of the short range
attraction and the long range repulsion.  Despite these recent
progresses, a deeper comprehension of the still puzzling
liquid-gel transition in colloidal system is requested.

Aqueous suspensions of charged platelike colloids have been the
subject of intense experimental and theoretical investigations.
The synthetic Hectorite clay Laponite is a perfect model for these
charged platelike colloids. It is in fact composed of nearly
monodisperse, rigid, disc-shaped platelets with a well defined
thickness of 1 nm, an average diameter of about 30 nm and a
negative surface charge of a few hundred $\emph{e}$. In spite of
intensive research on Laponite suspensions motivated by its
important industrial applications, there is no general agreement
about the mechanism of gelation; both attractive and repulsive
forces are claimed to be responsible for the gel formation
attributed by different authors to Wigner glass transition
\cite{Bonn}, frustrated nematic transition \cite{Mourchid, Kroon,
Gabriel, Mourchid1}, micro-segregation \cite{Mourchid, Pignon2,
Martin}, gelation \cite{Mongondry, Nicolai1, Nicolai2}, etc.
Hence, studies of the aggregation in Laponite suspensions will
give an important contribution to the understanding of the gel
formation process in systems where repulsive long range and
attractive short range interactions compete.

According to the phase diagram obtained from Mourchid \textit{et
al.} \cite{Mourchid} at ionic strength below $I=10^{-2}$ M
Laponite suspensions can be in two different physical states. Low
concentration suspensions ($C_w<C_w^*(I)$) form stable,
equilibrium fluid phase. Higher concentration suspensions ($C_w
\gtrsim C_w^*(I)$) are initially fluids but experience aging and
pass into a gel phase after a time that depends on the clay
amount. Several dynamical scattering studies have investigated the
aging process of samples belonging to this high $C_w$  phase
region \cite{Kroon1, Bonn, Knaebel, Abou, Bellour}, while only two
recent papers \cite{Nicolai1, Bellour} report about the lower
concentration region. While the first paper \cite{Nicolai1}
reports about the formation of a gel that does not flow when
tilted at low laponite concentration ($C_w$=1 $\%$, $I$= 5 mM
NaCl), the authors of the second paper \cite{Bellour} find that
samples at low concentrations ($C_w$=(0.1$\div$1) $\%$, $I\leq 5
\cdot 10^{-4} M$)are stable liquid suspensions. For this reason
and to try to understand the gelation mechanism for this charged
colloidal system we have performed accurate and systematic
measurements on several samples in the low concentration region
and at higher clay concentrations.

In this letter we present a dynamic light scattering study at
increasing concentrations from 0.3 to 3.1 wt $\%$ at ionic
strength between $I\simeq 10^{-4}$ M and $I\simeq 10^{-3}$ M where
$C_w^* \approx 1.5 \div 1.8 \%$ \cite{Mourchid}. A new gel region,
at $C_w\leq C_w^*$, has been identified and the gelation processes
in the two gel phases (at $C_w\leq C_w^*$ and at $C_w > C_w^*$)
have been accurately studied. Two different routes to gelation for
the two regions have been found and a possible explanation for the
processes involved in the building of the arrested phases has been
proposed.

Laponite is a hectorite synthetic clay manufactured by Laporte
Ltd, who kindly supplied us the material. Particular attention has
been devoted to samples preparation to avoid dissolution of
Laponite platelets that occurs if the samples are exposed to air
contamination due to the presence of atmospheric CO$_2$
\cite{Thompson, Mourchid2} and can seriously affect the
measurements. For this reason the whole procedure has been
performed in a glove box under N$_{2}$ flux and the samples have
been always kept in safe atmosphere during and after sample
preparation. The powder was firstly dried in an oven at $T$=400 K
for 4 hours (up to 20\% of the as received powder weight is due to
adsorbed water), then the powder is dispersed in deionized (pH=7)
water, stirred vigorously until the suspensions were cleared and
then filtered through 0.45 $\mu m$ pore size Millipore filters.
For same of the samples the pH has been measured by a Crison Glp
22 pHmeter, the final value, reached after 10$\div$20 min from the
laponite dispersion in water, was in the range pH=9.8$\div$10.0.
The final value of  pH=10, neglecting the contribution of Na
counterions released from some of the platelets to the solution,
would fix the ionic strength around $10^{-4} M$. Assuming that 50
counterions are released by each laponite platelet
\cite{Mongondry}, the ionic strength never exceeds $\sim 10^{-3}
M$ for all the investigated samples. The fact that laponite
dissolution does not take place in our samples is confirmed by NMR
spectroscopy \cite{Capuani} that checked continuously the amount
of Na and Mg ions present in the solutions. These measurements
show in fact that, both in the low and in the high concentration
ranges, the fraction of solvated Na ions does not change with
time, and that the amount of solvated Mg ions are below the
instrumental sensibility. These observations indicate that the
chemical reaction of Laponite dissolution \cite{Thompson,
Mourchid2} is not affecting our samples. Futhermore, in none of
the samples studied we observed any evidence of sedimentation also
several months after their preparation as has been checked by
direct sample inspection and by performing test measurements of
the correlation function at different heights in the cells. The
starting aging time (t$_w$=0) is defined as the time when the
suspension is filtered. This sample preparation procedure is
similar, but not identical, to that already used in previous works
\cite{Nicolai1, Nicolai2,Abou, Bellour}. Some authors do not
report information on the drying of the powder, others allow the
suspension to get in contact with atmosphere, and different
details can lead to small differences in the sample behavior on
aging. In addition, the determination of the sample concentration
is also problematic, as the filtering procedure can alter the
actual concentration if large clusters are present in the
suspension after the stirring phase. To minimize this effect, we
have prepared a large amount of concentrated solution, then -after
the filtering- we have diluted this parent sample at the desired
concentrations. In this way, in the present measurements, the {\it
relative} concentrations are well defined, even if the absolute
value in $C_w$ can be affected by a systematic error, that we
estimate in the range $\Delta C_w/C_w \approx$ 0.1.

Dynamic light scattering measurements were performed using an
ALV-5000 logarithmic correlator in combination with a standard
optical set-up based on a He-Ne ($\lambda$ = 632.8 nm) 10 mW laser
and a photomultiplier detector. The intensity correlation function
was directly obtained as $g_2(q,t)=<I(q,t)I(q,0)>/<I(q,0)>^{2}$,
where $q$ is the modulus of the scattering wave vector defined as
$q=(4\pi n/\lambda)sin(\theta/2)$ ($\theta$=90$^o$ in the present
experiment). For the low concentration samples ($C_w\leq 1.5 \%$)
the measurements have been repeated once a week for a long period
of time (up to four months for the lowest concentration), while
for $C_w>1.5 \%$ the sample was left in the scattering system and
photocorrelation spectra were continuously acquired until the
gelation was reached.

\begin{figure}[h]
\centering
\includegraphics[width=.43\textwidth]{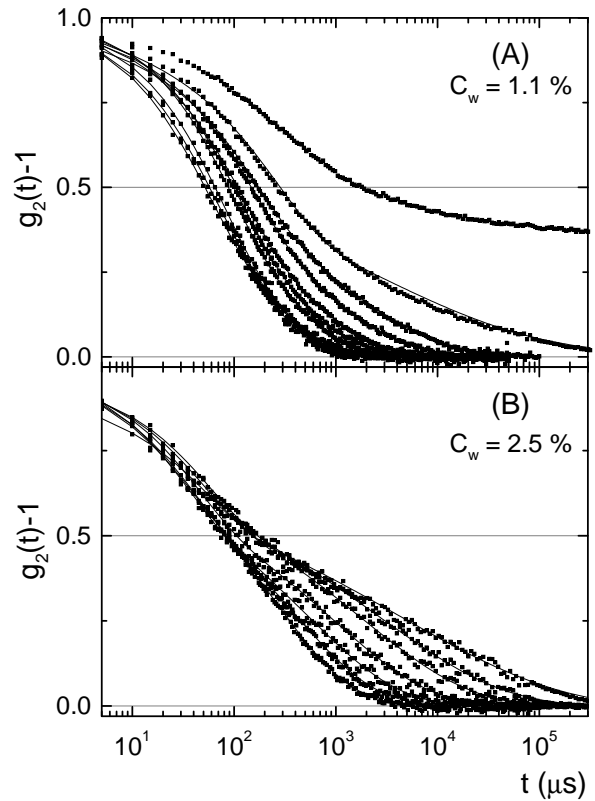}
\caption{ Evolution of the measured intensity correlation
functions (symbols) and corresponding fits
  with Eq.~(\ref{fit}) (continuous lines) for two different
  Laponite suspensions at the indicate concentrations
   at different waiting times t$_w$. The curves are measured at
   increasing waiting times that (from left to right)
   are t$_w$=288, 792, 1128, 1464, 1632, 1800, 1968, 2136, 2328, 2640
hours for sample (A) and t$_w$=6, 30, 54, 78, 102, 126, 150 hours
for sample (B)(for which only data in the ergodic phase are
reported).} \label{f1}
\end{figure}

As an example, correlation functions at increasing aging times
t$_w$ for two different samples at low- (Fig. 1A - $C_w=1.1$ wt
$\% <C_w^*$ ) and high- concentration (Fig. 1B $C_w=2.5$ wt $\%
>C_w^*$), are reported in Figure~1. As it is evident from the figure
both the samples are doing aging, the dynamics is in fact becoming
slow and slow for increasing waiting time t$_w$. This behavior was
expected for the higher concentration sample of Fig.~1B, several
measurements \cite{Kroon1,Bonn,Knaebel,Abou,Bellour}, indeed, have
shown the existence of aging and a sol-gel transition for  long
enough waiting times. The aging process was instead
 unexpected for the lower concentration sample of
Fig.~1A where the liquid state is predicted from the phase diagram
\cite{Mourchid}, and confirmed by recent dynamic light scattering
measurements \cite{Bellour} where, perhaps, the authors did not
wait for a time long enough to observe significant aging. From
Fig.~1A it is instead evident that the system is aging and for the
longest waiting time reported (t$_w$=2640 hours) there is a
qualitative change in the correlation function: a crossover
between a complete and an incomplete decay. This behavior is the
indication of a strong ergodicity breaking, signature of a sol-gel
transition, as already observed in \cite{Kroon1, Bonn} for samples
in the higher concentration region and in \cite{Nicolai1} for a
sample at $C_w=1 \%$ with $I=5\cdot10^{-3} M$. Our dynamic light
scattering measurements show that for all the samples studied,
down to the lowest concentrated one  ($C_w=0.3 \%$), there is a
typical waiting time, $t_{w}^{\infty}$, increasing with decreasing
clay concentration $C_w$, at which the system undergoes the
gelation. Depending from the initial concentration this gelation
time can vary from hours to several months.

Figure~1 shows also that the correlation functions decay following
a two steps behavior, i.~e. there are two different relaxation
processes, a fast and a slow ones. For this reason the fitting
expression should contain two contributions. In this case the
squared sum of an exponential and a stretched exponential function
is used, as already reported in Ref.\cite{Abou}:

\begin{equation}
g_2(q,t)-1=b \left ( a e^{(-t/\tau_1)}+(1-a)
e^{(-(t/\tau_2)^\beta)} \right )^2 \label{fit}
\end{equation}

where $b$ represents the coherence factor. The stretched
exponential, instead than another simple exponential term, is used
since it has been found to give a good description of the slow
relaxation process in glassy systems. The fitting expression well
describe the photocorrelation spectra for all the aging times in
the liquid (ergodic) phase and for all the investigated
concentrations with the relaxation time $\tau_1$ associated to the
fast dynamics and the relaxation time $\tau_2$ and the stretching
parameter $\beta$ that describe the slow part of the
autocorrelation function. The fits are shown as full lines in
Fig.~1 and the resulting chi square is always within its standard
deviation.

Another important peculiarity of the correlation functions, that
can be directly observed in the raw data reported in Fig.~1, is
that the aging process evolves differently for the lower and
higher concentration samples. While in fact the initial decay of
the correlation functions -fast dynamics, characterized by the
correlation time $\tau_1$- remains constant with waiting time for
the higher concentration sample (Fig. 1B) it seems to become
slower with increasing waiting time for the lower concentration
sample (Fig.~1A). This different behavior is confirmed from the
results of the fits, that also show another major difference
between the samples at low and high concentrations.

In the following we will report the $t_w$ and $C_w$ dependence of
the fitting parameters, in particular we focus on the behavior of
the parameters of the slow decay: the relaxation time $\tau_2$,
the stretching exponent $\beta$ and the "mean" relaxation time
$\tau_m$:

\begin{equation}
 \tau_m=\tau_2  \frac{1}{\beta} \Gamma(\frac{1}{\beta}).
 \label{taumeq}
\end{equation}

where $\Gamma$ is the usual Euler gamma function.

\begin{figure}[h]
\centering
\includegraphics[width=.43\textwidth]{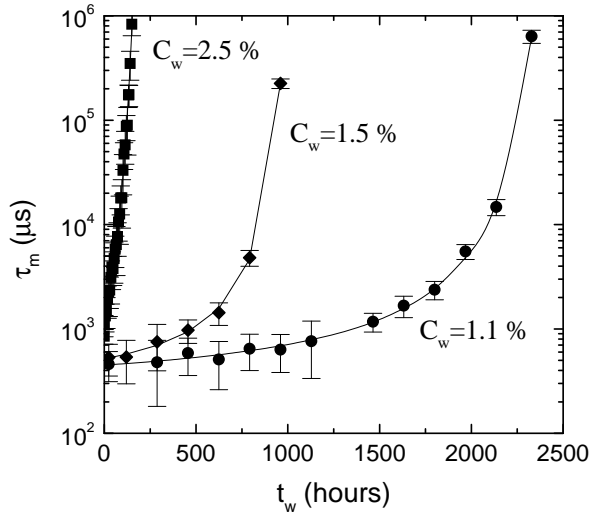}
\caption{Waiting time dependence of $\tau_m$ (see
Eq.~\ref{taumeq}), i.~e. of the average relaxation time of the
slow decay of the correlation functions as those reported in
Fig.~\ref{f1}. As an example, the $t_w$ dependence of $\tau_m$ is
reported for the three indicated concentrations. Continuous lines
are fits of the data with Eq.~\ref{taum}.} \label{f2}.
\end{figure}

As an example, the values of $\tau_m$ obtained for three different
concentrations are reported in Fig.~2. It is evident the common
behavior of $\tau_m$ that seems to diverge at a given
$t_w=t_{w}^{\infty}$ i.e. when the gelation occurs. This clearly
shows that, as already seen from the raw correlation functions,
also the samples at concentrations lower/equal to 1.5 wt $\%$ are
actually doing aging and undergo a gelation transition. To have
more information about this gelation process we represent the
aging time ($t_w$) dependence of mean relaxation time ($\tau_m$)
with the law:

\begin{figure}[h]
\centering
\includegraphics[width=.43\textwidth]{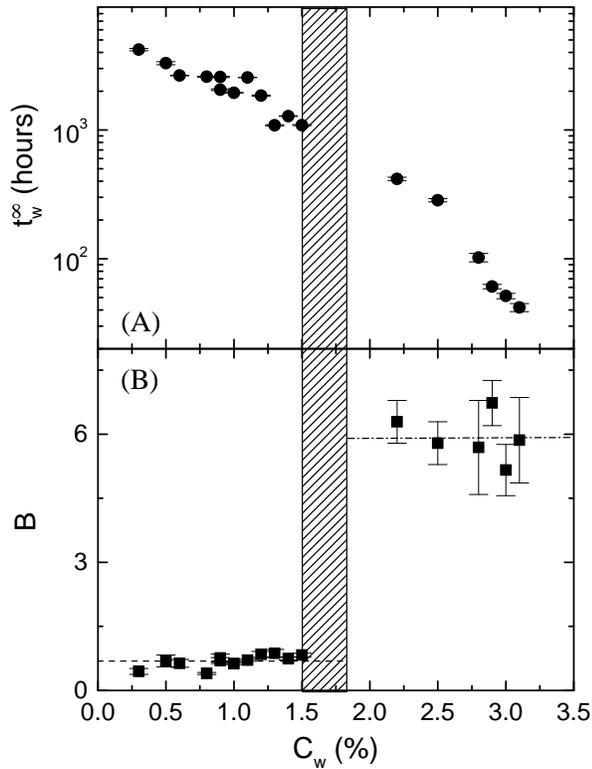}
\caption{Concentration dependence of the divergence time
$t_{w}^{\infty}$ (A) and of the $B$ parameter (B), entering in
Eq.~\ref{taum}. The dashed lines are guides to the eyes. The
shadow area indicates the region where the liquid-gel transition
was supposed to be according to Ref. \cite{Mourchid}} \label{f3}
\end{figure}

\begin{equation}
 \tau_m = \tau_0 \; exp \left (B \frac{t_w}{t_{w}^{\infty}-t_w}
 \right ) \label{taum}
\end{equation}

We do not have any rationale behind such a law, it can be
considered a generalization to long waiting time of the
exponential growth with $t_w$ observed for $\tau_2$ by Abou
\textit{et al.} \cite{Abou} in the high concentration samples.
Equation~\ref{taum} -reported in Fig.~2 as full lines- well
describe the measurements. The most significative parameters of
the fits, $t_{w}^{\infty}$ and $B$, for all the studied
concentrations are shown in Fig.~3. Here the vertical dotted
region indicates the range of concentrations that, in the phase
diagram \cite{Mourchid}, would mark the transition from the
"liquid" to the gel phase in the range of $I\simeq 10^{-3} -
10^{-4} M$.  The results of the fit indicate that $t_{w}^{\infty}$
-that can be considered as the time at which the gelation actually
occurs- is continuously decreasing with increasing clay
concentration, without any evident discontinuity in correspondence
of the "transition" region. The concentration dependence of the
parameter $B$, which measures how fast $\tau_m$ approaches the
divergence, is shown in Fig.~3B. This parameter is almost constant
for all the samples in the low concentration region, while shows a
large discontinuity on passing in the higher concentration region.
It is important to note that this jump takes place in a region
that encompass the supposed "liquid-gel transition" region.

\begin{figure}[h]
\centering
\includegraphics[width=.43\textwidth]{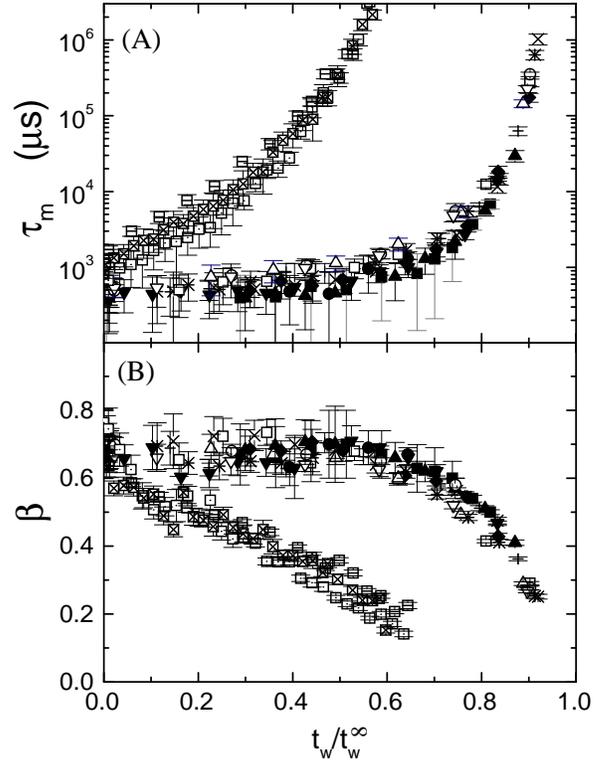}
\caption{The waiting time dependence of the $\tau_m$ (A) and
$\beta$ (B) are reported as a function of the scaled variable
$t_w/t_{w}^{\infty}$ for some of the investigated concentrations:
$\blacksquare = 0.3 \%$, $\bullet = 0.5 \%$, $\blacktriangle = 0.6
\%$, $\blacktriangledown = 0.8 \%$, $\blacklozenge = 0.9 \%$, + $=
0.9 \%$, $\times = 1.0 \%$, $\ast = 1.1 \%$, $\square = 1.2 \%$,
$\circ = 1.3 \%$, $\triangle = 1.4 \%$, $\triangledown = 1.5 \%$,
$\boxdot = 2.2 \%$, $\boxtimes = 2.5 \%$, $\boxminus = 2.8 \%$.
The data for $C_w \leq 1.5 \%$ collapse on a single master curve.}
\label{f4}
\end{figure}

The fact that the value of the $B$ parameter is almost constant
for all the low concentration samples is an indication of the
existence of a scaling law. Indeed, if we plot the quantities
$\tau_m(t_w)$ and $\beta(t_w)$ as a function of
$t_w/t_{w}^{\infty}$ all the different $C_w$ data with almost
constant $B$ should collapse on a single master curve. The
$\tau_{m}$ and the $\beta$ parameters in function of the
normalized waiting time are reported in Fig.~4A and 4B
respectively. As expected, all the data for both the $\tau_{m}$
and the $\beta$ parameters of lower concentrations collapse,
within their statistical uncertainties, on a single curve while
the data at higher clay concentrations have a different behavior.
We already observed from the direct comparison of Fig.~1A with
Fig.~1B that the aging process is qualitatively different for the
low and high concentration samples. Figures~3 and 4 quantify this
difference in the physical properties characterizing the aging
phenomenon in the two different concentration regions.

In conclusion the present observations indicate that the stable
phase of Laponite suspensions in pure water at $C_w \lesssim
C_w^*$ -expected to be liquid according to previous studies
\cite{Mourchid}- actually is an arrested phase. Recently Nicolai
\textit{et al.} \cite{Mongondry} proposed a revisitation of the
phase diagram and suggested that the equilibrium state in the low
concentration region is not liquid but a very fragile gel that
takes a long time to appear \cite{Nicolai2}. Probably this "long
time" is the reason why previous measurements indicated this phase
as liquid. Our measurements clearly show not only that at all the
investigated concentrations the stable phase is an arrested phase
but also that the aggregation process is basically different for
the samples at $C_w \lesssim C_w^*$ and those at $C_w \gtrsim
C_w^*$. This is evident from the behavior of $\tau_m$ and $\beta$
reported in Fig.~4 and from the discontinuity of the $B$ parameter
across the $C_w = C_w^*$ region as shown in Fig.~3. In this sense
the "liquid-gel" transition line of the phase diagram
\cite{Mourchid} is not a real liquid-gel transition but rather
seems to indicate a sort of "fragile gel-gel" transition.

The origin of the different routes towards an arrested phase
observed for low- and high-concentration Laponite suspensions
calls for an explanation that goes beyond the aim of the present
work. At the level of speculation, it is worth to recall that the
microscopic interaction between Laponite plates is due to a
screened Coulomb interaction, which can be modeled by a
Yukawa-like repulsion at long distances and a quadrupolar electric
and/or a van der Waals terms at short distances \cite{Dijkstra,
Trizac}. This competition of short-range attraction and long range
repulsion resembles that recently proposed \cite{Sciortino} to
describe the phenomenology of colloidal gels. In
Ref.~\cite{Sciortino} it has been suggested that the gelation
taking place in attractive colloidal suspensions at very low
concentration involves the growth of larger and larger clusters
(driven by the short range attraction). The glass transition of
these clusters, driven by the long range repulsion, is responsible
for the arrest of the dynamics and the formation of a gel phase.
The authors prove that the gel is essentially a Wigner glass,
composed by clusters of colloidal particles. We can speculate that
low concentrations Laponite suspensions behave as the attractive
colloids in Ref.~\cite{Sciortino}. This scenario is in agreement
with the fact that the long time needed to form the arrested phase
in the low concentration range is spent by the system to build up
the clusters. Also at high Laponite concentration the arrested
state could be a Wigner glass but in this case the packing
fraction of the Debye Huckel sphere associated to each platelets
reaches values as high as 0.43 \cite{Bonn} and the single laponite
platelet would be the elementary constituent of the Wigner glass.

\end{document}